\documentclass[
    aps,
    pra,
    twocolumn,
    letterpaper,
    10pt,
    superscriptaddress,
    notitlepage,
    amsmath,
    amssymb,
    floatfix
]{revtex4-1} 
 
\usepackage{graphicx}
\usepackage{dcolumn}
\usepackage{bm}
\usepackage{color}
\usepackage{amssymb}
\usepackage{amsmath}
\usepackage{hyperref}
\usepackage[normalem]{ulem}

\newcommand{\bel}{\begin{equation}}
\newcommand{\eel}{\end{equation}}
\newcommand{\be}{\begin{equation*}}
\newcommand{\ee}{\end{equation*}}
\newcommand{\bal}{\begin{eqnarray}}
\newcommand{\eal}{\end{eqnarray}}
\newcommand{\ba}{\begin{eqnarray*}}
\newcommand{\ea}{\end{eqnarray*}}

\newcommand{\refeq}[1]{Eq.~(\ref{#1})}
\newcommand{\reffig}[1]{Fig.~\ref{#1}}

\newcommand{\ev}[1]{\langle #1 \rangle}
\newcommand{\ket}[1]{| #1 \rangle}
\newcommand{\bra}[1]{\langle #1 |}

\newcommand{\s}{^\ast}

\newcommand{\PP}{\mathcal{P}}

\renewcommand{\d}[1]{\!d#1\,}

\begin{document}

\title{A quantum network of clocks}

\author{P. K\'{o}m\'{a}r}
\thanks{These authors contributed equally to this work}
\affiliation{Physics Department, Harvard University, Cambridge,
Massachusetts 02138, USA}

\author{E. M. Kessler}
\thanks{These authors contributed equally to this work}
\affiliation{Physics Department, Harvard University, Cambridge,
Massachusetts 02138, USA}
\affiliation{ITAMP, Harvard-Smithsonian Center for Astrophysics, Cambridge, MA
02138, USA}

\author{M. Bishof}
\affiliation{JILA, NIST, Department of Physics,  University of Colorado,
Boulder, CO 80309, USA}

\author{L. Jiang}
\affiliation{Department of Applied Physics, Yale University New Haven, CT
06520, USA}

\author{A. S. S{\o}rensen}
\affiliation{QUANTOP, Danish National Research Foundation Centre of Quantum Optics, Niels Bohr Institute,
DK-2100 Copenhagen, Denmark}

\author{J. Ye}
\affiliation{JILA, NIST, Department of Physics,  University of Colorado,
Boulder, CO 80309, USA}

\author{M. D. Lukin}
\affiliation{Physics Department, Harvard University, Cambridge,
Massachusetts 02138, USA}

\date{\today}

%


\maketitle


\textbf{The development of precise atomic clocks has led to many scientific and
technological advances that play an increasingly important role in  modern
society. Shared timing information constitutes a key resource for  positioning
and navigation with a direct correspondence between timing accuracy and
precision in applications such as the Global Positioning System (GPS).
By combining  precision metrology and quantum networks, we propose here a
quantum, cooperative protocol for the operation of a network consisting of
geographically remote optical atomic clocks. Using non-local entangled states,
we demonstrate an optimal utilization of the global network resources, and show
that such a network can be operated near the fundamental limit set by quantum
theory yielding an ultra-precise clock signal.
Furthermore, the internal structure of the network, combined with basic
techniques from quantum communication, guarantees security both from internal
and external threats. Realization of such a global quantum network of clocks may
allow construction of a real-time single international time scale (world clock)
with unprecedented stability and accuracy.}

\begin{figure}[!]
\centering
\includegraphics[width=0.49\textwidth]{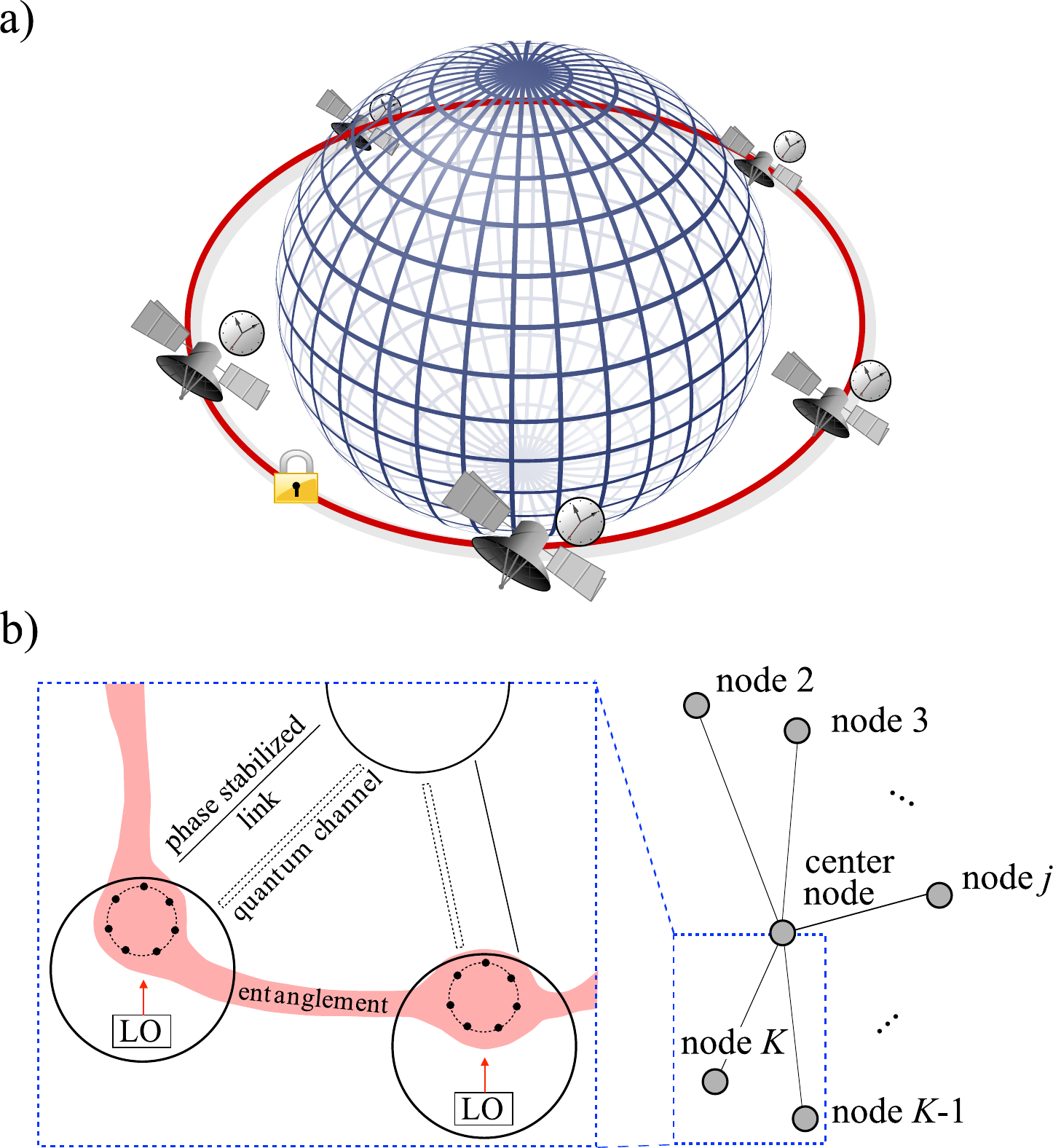}
\caption{ 
\label{fig:1} The concept of world-wide quantum clock network. 
a) Illustration of a cooperative clock operation protocol in which individual parties (e.g., satellite based atomic clocks from different countries) jointly allocate
their respective resources in a global network involving entangled quantum states. This guarantees an
optimal use of the global resources, achieving an ultra-precise clock signal
limited only by the fundamental bounds of quantum metrology and, 
in addition, guaranteeing secure distribution of the clock signal.  
 b) In addition to locally operating the individual clocks, the different nodes
 (i.e., satellites) employ network-wide entangled states to interrogate their
 respective local oscillators (LOs). The acquired information is sent to a particular node serving as
 a center where it is used to stabilize a center of mass mode of the different
 LOs. This yields an ultra-precise clock signal accessible to all network
 members. }
\end{figure}
 
With the advances of highly phase coherent lasers, optical atomic clocks
containing multiple atoms have demonstrated stability that reaches the standard
quantum limit (SQL) set by the available atom number within a clock
\cite{Nicholson2012, Hinkley2013}.
Reaching beyond SQL, we stand to gain a significant improvement of clock
performance by preparing atoms in quantum correlated states (e.g., spin
squeezed states \cite{Leroux2010}). Here we describe a new approach to maximize
the performance of a network composed of multiple clocks allowing to gain
advantage of all rescources available at each node.
Several recent advances in precision metrology and quantum science make this approach realistic.  On the
 one hand, capabilities to maintain phase coherent
optical links spanning the entire visible spectrum and over macroscopic
distances have been demonstrated, with the capability of delivering the most
stable optical oscillator from one color or location to another \cite{Ye2003,
Droste2013}. 
On the other hand, quantum communications and entanglement techniques are
enabling distant quantum objects to be connected in a quantum network
\cite{Cirac1997, Kimble2008, Perseguers2013}, that can enable novel,
extraordinary capabilities.
Combining these two technological frontiers, we show here that  a distributed
network composed of quantum-limited clocks separated by large distances -- as
appropriate, e.g., for the satellite-based clocks possibly operated by different
nations -- can be operated  as an ultimate ``world clock'', where all members
combine their individual resources in a quantum coherent way  to achieve greater
clock stability and distribute this international time scale in real time for
all. 

The distributed achitecture allows each participant of the network to profit
from a stability of the local clock signal that is enhanced by a factor
proportional to the total number of parties (as compared to an independent
operation of the individual clocks) without losing sovereignty or compromising
security. This cooperative gain strongly incentivizes joining the collaborative
network while retaining robustness against
 disruptions of communication channels by allowing the parties to fall back to
 individual clock operation.
Our scheme is superior to an alternative approach of disseminating the time
signal from a single location containing all qubits, since errors arising from
imperfect phase links can be largely reduced by relying on the stabilized
and locally available local oscillators.
We demonstrate that by preparing quantum-correlated states of
remote clocks, the network can yield the best possible clock signal allowed by
quantum theory for the combined resources.
Furthermore, enabled through the use of quantum communication techniques,  such
a network can be made secure, such that only parties contributing to its
operation may enjoy the benefit of an ultra-precise clock signal. Besides
serving as a real-time clock for the international time scale,  the proposed
quantum network also represents a large-scale quantum sensor that can be used to
probe the fundamental laws of physics, including relativity and connections between space-time and quantum physics.


\section*{The concept of quantum clock network}
\label{sec:QCN}

 \reffig{fig:1} illustrates the basic concept for the proposed quantum
clock network.
We consider a set of $K$ atomic clocks (constituting the nodes of the network), each based on a large number of atoms (clock
qubits) serving as the frequency reference $\omega_0$ at different geographical
locations. In our approach,  each clock has its own independently operated local
oscillator (LO), $\mathcal{E}_j(t)\propto e^{i\nu_j t}$, with detuning $\delta_j
= \nu_j - \omega_0$, $(j=1,2\dots K)$. It keeps the time by interrogating its
qubits periodically, and uses the measurement data to stabilize the LO frequency
at the reference frequency of the atomic transition. However, as opposed to the
conventional approach, in which each LO interrogates its own independent qubits,
we consider the situation in which each network node  allocates some of its
qubits to form entangled states stretching across all nodes. When interrogated
within a properly designed measurement scheme, such entangled network states
provide ultra-precise information about the deviation of the center-of-mass
(COM) frequency $\nu_\textrm{COM} = \sum_j \nu_j/K$ of all local oscillators
from the atomic resonance.



Each clock cycle consists from three stages: preparation of the clock atom state
(initialization), interrogation by the LOs (measurement) and correction of the
laser frequency according to the measurement outcome (feedback).
In the further analysis, we assume, for convenience, that in each interrogation
cycle one of the nodes plays the role of an alternating center, which initiates
each Ramsey cycle and collects the measurement data from the other nodes via classical
channels [\reffig{fig:1}~b)], as well as LO signals via optical links,  to
feedback the COM signal.  (In practice, it is straightforward to devise a
similar network with same functionality and a flat hierarchical structure where
no center is needed, see Supplementary Information).  This information, in turn,
can be utilized in a feedback cycle to yield a Heisenberg-limited stability of
the COM clock signal generated by the network, which is  subsequently
distributed to the individual nodes in a secure fashion.  As a result, after a
few cycles, the LOs corresponding to each individual node achieve an accuracy
and stability effectively resulting from interrogating atoms in the entire
network.

\section*{Preparation of network-wide entangled states}
\label{sec:NWES}

In the initialization stage of each clock cycle, entangled states spanning
across the nodes at different geographical positions of the network are
prepared. In the following, we describe exemplarily how a single network-wide
GHZ state can be prepared. 
The entangled states employed in the proposed quantum network protocol -- which
are described in the following section -- consist of products of GHZ states of
different size. They can be prepared by repetition of the protocol that
we now describe.
 
\begin{figure}
\centering
\includegraphics[width=0.48\textwidth]{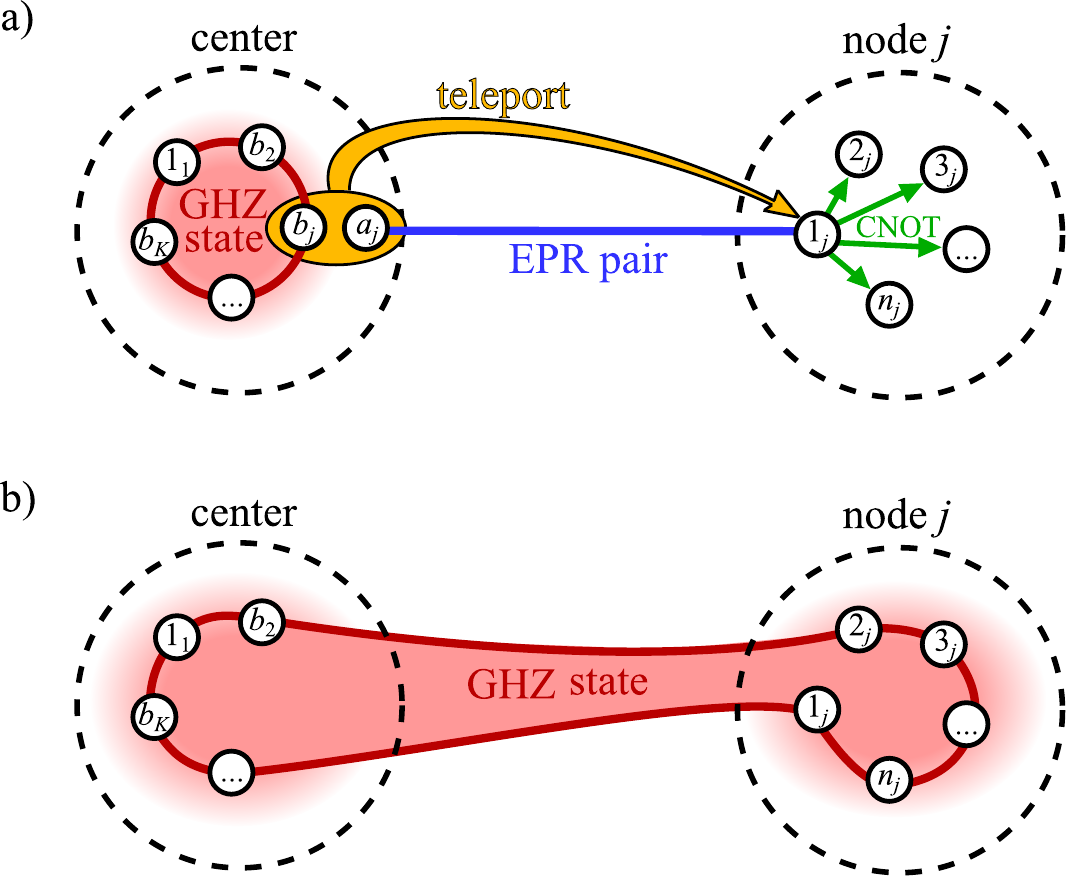}
\caption{ 
\label{fig:entangling}
Entangled state preparation between distant nodes. 
a) The center node $(j=1)$ initiates the initialization sequence by preparing a
local GHZ state accross the qubits $\{b_j\}_{j=2}^K$ and $1_1$, as well as $(K-1)$ EPR pairs on
the qubit pairs $\{(a_j,1_j)\}_{j=2}^K$.
Quantum teleportation expands this GHZ state to the first qubit within each of
the individual nodes. 
b) Originating from the teleported qubits, the nodes grow the GHZ state to
involve all the desired local qubits by employing local entangling operations.
The procedure results in a common GHZ states over all atoms of the nodes.
 }
\end{figure}

For simplicity, we assume that each node $j$ ($j=1,\dots K$) contains an
identical number $n$ of clock qubits which we label as $1_j, 2_j,\dots n_j$ (in
the Supplementary Information we discuss the case where the nodes contain
different amounts of clock qubits).
Further, we assume, for convenience, that the center node ($j=1$) has access to
additional $2(K-1)$ ancilla qubits $a_2,\dots, a_K,b_2,\dots,b_K$ besides the
$n$ clock atoms (a slightly more complicated procedure allows to refrain from
the use of ancilla qubits, see Supplementary Information).
The entangling procedure starts at the center with the creation of a fully
entangled state of one half of the ancilla qubits $\{b_j\}$, and its first clock
qubit $1_1$. This can be realized, e.g.  with a single qubit $\pi/2$-rotation
(on qubit $1_1$) and a series of controlled not (CNOT) gates
\cite{Nielsen:2000vn} (between $1_1$ and each $b_j$).
The result is a GHZ state, $[\ket{00\dots 0}_{1_1,b_2,b_3,\dots b_K} +
i\ket{11\dots 1}_{1_1,b_2,b_3,\dots b_K}]/\sqrt{2}$.
In parallel, the center uses the other half of the ancillas $\{a_j\}$ to create single EPR pairs with each node $j\neq1$,
either by directly sending flying qubits and converting them to stationary
qubits, or by using quantum repeater techniques to prepare high-fidelity entanglement \cite{Duan2001}. As a result of this procedure, 
one part of the pair is stored at the center node (qubit $a_j$), while the other one
is stored at the $j$th  node (qubit $1_j$), forming the states $[\ket{00}_{a_j,1_j} +
\ket{11}_{a_j,1_j}]/\sqrt{2}$ for every $j$ (see \reffig{fig:entangling}).

%

Next, the center performs $K-1$ separate Bell measurements on
 its ancilla qubit pairs $\{(b_j, a_j)\}$. This teleports the state of qubit $b_j$ to
 qubit $1_j$
($j=2,\dots K$), up to a local single-qubit rotation, which is performed
after the measurement outcomes are sent to the node via classical channels.
The result of the  teleportations is a collective GHZ state
$\frac{1}{\sqrt{2}}\ket{00\dots 0}_{1_1,1_2,\dots 1_K} + i \ket{11\dots
1}_{1_1,1_2,\dots 1_K}$, stretching across the first qubits of all $K$ nodes.

In the final step of entangling, all nodes (including the center) extend the
entanglement to all of their remaining clock qubits. To do this, each node $j$
performs a series of CNOT gates controlled on $1_j$ and targeting qubits $2_j,
3_j, \dots n_j$.  At the end of the protocol the different nodes share a common
GHZ state $[\ket{\mathbf{0}} + i\ket{\mathbf{1}}]/\sqrt{2}$, where
$\ket{\mathbf{0}}$ and $\ket{\mathbf{1}}$ are product states of all qubits
$\{i_j\;:\; i=1,2,\dots n,\; j=1,2,\dots K\}$ being in $\ket{0}$ or $\ket{1}$,
respectively. As discussed below, in practice the entanglement distribution can
be done  either via polarization- or frequency-entangled photons with frequency
difference in the microwave domain, in which case the ancillary qubits involved
in the entanglement distribution will be different from the clock qubits.
Typically, as part of the preparation process, time delays arise between the
initialization of different clock qubits. Its detrimental effects can be
entirely avoided by proper local timing or prior preparation of entanglement, as
discussed in the Supplementary Information.

\section*{Interrogation}
\label{sec:SA}

The use of entangled resources (in form of network-wide GHZ-like states) during the interrogation phase enables an optimal use of the available resources via the following procedure.  Assume we have a total of $\tilde N$ 
qubits at
our disposal which are equally distributed between the $K$ nodes (indexed
$j=1,\hdots K$) and prepared in a non-local GHZ state $[\ket{\mathbf{0}} +
i\ket{\mathbf{1}}]/\sqrt{2}$, where $\ket{\mathbf{0 (1)}}\equiv\ket{0(1)}^{\otimes
\tilde N}$. During the
interrogation time $T$, a clock qubit at node $j$ picks up a relative phase $\phi_j = \delta_j
T$.
Due to the non-local character of the state, these phases accumulate in the total
state of the atoms  $[\ket{\mathbf{0}} + i e^{i\Phi}\ket{\mathbf{1}}]/\sqrt{2}$,
where the collective phase after the interrogation time $T$ is given as
\bel
\label{eq:1}
\Phi = \sum_{j=1}^K \frac{\tilde N}{K} \phi_j =
\tilde N \delta_\text{COM} T,
\eel
where $\delta_\text{COM} = \nu_\text{COM} - \omega_0$.
To extract the phase information picked up by the different GHZ states, after
each interrogation phase, the individual nodes $j$ measure their respective
qubits in the $x$-basis, and evaluate the parity of all measurement outcomes
$p_j$.
Subsequently, the nodes send this information to the center node via a classical
channel, where the total parity $p = \prod_{j} p_{j}$ is evaluated, and the
phase information is extracted \cite{Bollinger1996, Leibfried2004}. Note, that
only the full set $\{p_j |j=1\hdots K \}$ contains information. This can be
interpreted as only the center node holding the key,  namely its own measurement
outcome $p_1$, to decode the phase information sent from the nodes.

The proportionality with $\tilde N$ in \refeq{eq:1} represents the quantum
enhancement in the estimation of $\delta_\text{COM}$. However, for realistic
laser noise spectra, this suggested enhancement is corrupted  by the increase of
uncontrolled phase slips for a single GHZ
state 
\cite{Wineland1998}: Whenever after the Ramsey time the phase $\Phi$ -- which
due to the laser frequency fluctuations constitutes a random variable itself --
falls out of the interval $[-\pi,\pi]$ the estimation fails. This limitation
restricts the maximal Ramsey time to values $T < (\tilde
N\gamma_\text{LO})^{-1}$, preventing any quantum gain in the estimation.

To circumvent this problem, we use entangled states consisting of products of
successively larger GHZ ensembles, see SI and 
\cite{Kessler2013}. In this approach, interrogated network atoms are split into
several independent, shared groups.
We write the number of the first group of atoms as $\tilde N =2^{M-1} K$, for
some natural number $M$.
Furthermore, the network shares additional groups of atoms, each containing
$2^{i} K$ ($i=0, \hdots M-2$) equally distributed between the nodes and prepared
in GHZ states. Finally, each node has a small number of uncorrelated atoms
interrogated by LOs. Using a protocol reminiscent of the phase estimation
algorithm \cite{Kessler2013,Nielsen:2000vn, Giedke2006} these states allow to
directly assess the bits $Z_i \in \{0,1\}$ of the binary fraction representation
of the laser phase $\Phi_\text{LO}=\delta_\text{COM} T =2\pi [(Z_1-1) 2^{-1} +
Z_2 2^{-2} + Z_3 2^{-3} \hdots]$. This
yields an estimate of $\Phi_\text{LO}$ with Heisenberg-limited accuracy, up to a
logarithmic correction, see SI:
\bel 
	\Delta \Phi_\text{LO} =\frac{8}{\pi}\log(N)/N,
	\label{eq2}
\eel 
even for Ramsey times beyond the limits of the laser frequency fluctuations [$T
> (\tilde N\gamma_\text{LO}^{-1})$], where $N$ represent the total number of
clock atoms employed in the scheme. The logarithmic correction arises due to the
number of particles required to realize this (incoherent) version of the phase
estimation algorithm.

 \section*{Feedback}

The measured value of the phase  $\Phi_\text{LO}$,
gives an estimate on the COM detuning
$\tilde\delta_\text{COM}$ after each Ramsey cycle, which is subsequently used
by the center node to stabilize the COM laser signal.
To this end the center generates the COM of the frequencies. Every node sends
its local oscillator field $\mathcal{E}_i$ to the center via phase-stable
optical links, and the center synthesizes the COM frequency $\nu_\text{COM}$ by
averaging the $\nu_j$ frequencies with equal weights \footnote{With different
number of qubits at each node, the weighted average needs to be taken.}.
This can be implemented via heterodyne beat of the local oscillator in the
center against each incoming laser signal, resulting in $K$ beat frequencies.
Synthesizing  these beat frequencies allows the
local oscillator of the central node to phase track $\nu_\text{COM}$.
The center distributes the stabilized clock signal to different members of the network by sending individual error
signals $\tilde \delta_j = \tilde \delta_\text{COM} + (\nu_j - \nu_\text{COM})$
to all nodes $j$, respectively, and corrects its own LO as well, accordingly.
Alternatively, the center can be operated to provide restricted feedback
information to the nodes, see SI.



\section*{Stability analysis}
\label{sec:comp}

In this section, we demonstrate that the proposed quantum clock network achieves
the best clock signal, allowed by quantum theory for the available resources, i.e.
the total atom number. Rather than individually operating their respective LOs the
joint use of resources allows the network to  directly interrogate and stabilize
the COM mode of the lasers. To quantify this cooperative gain, we compare
networks of different types (classical or quantum mechanical interrogation of
the respective LOs) and degrees of cooperation (no cooperation, classical, or
quantum cooperation).

\begin{figure}
\centering
\includegraphics[width=0.48\textwidth]{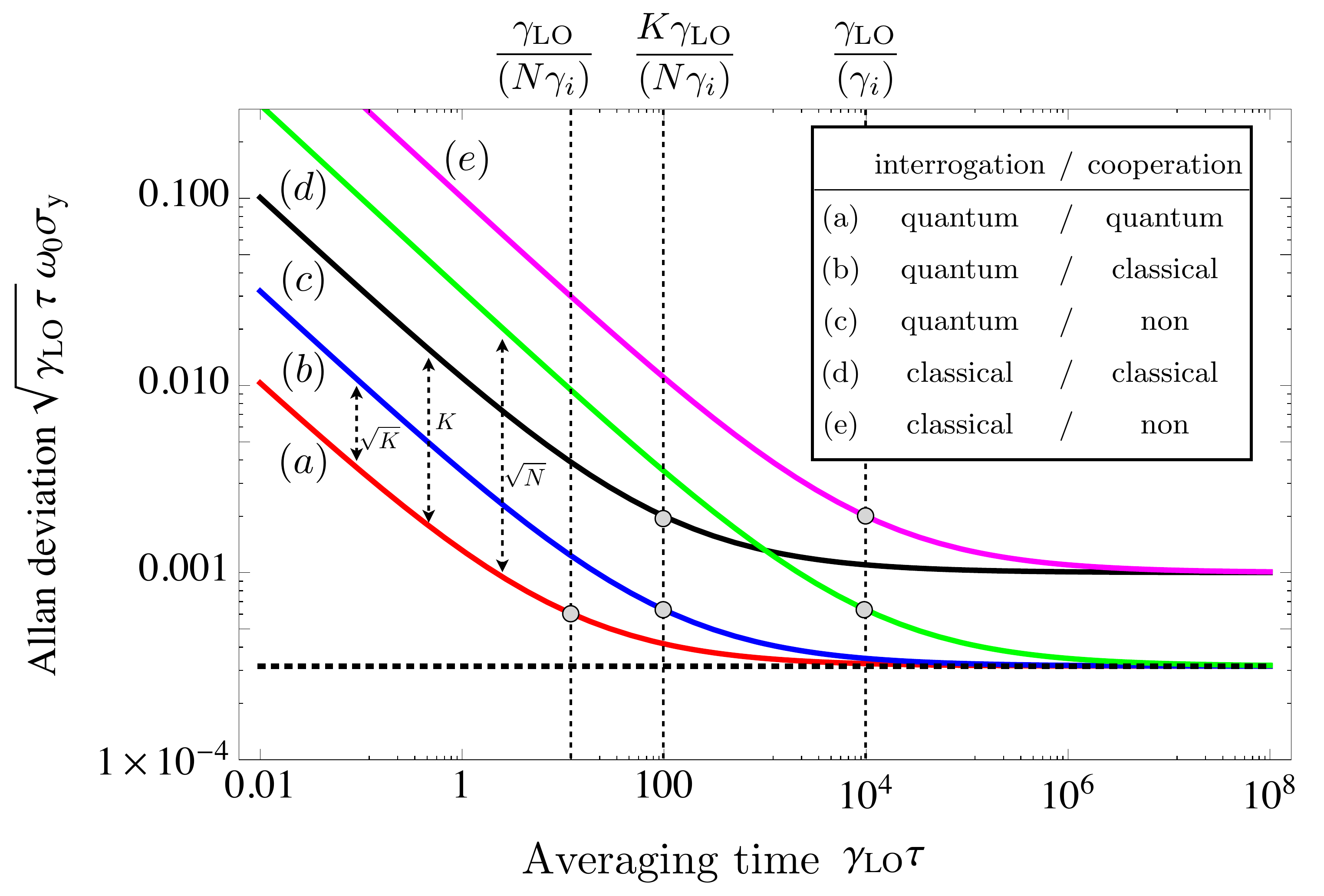}
\caption{ 
\label{fig:comp}
Performance of different operation schemes. Comparison of
the achievable (rescaled) Allan deviation $\sqrt{\gamma_\text{LO} \tau} \omega_0 \sigma_y$ using clock networks of different types and degrees of
cooperation. (a) the proposed protocol realizing quantum interrogation and
cooperation, (b) quantum interrogation and classical cooperation, (c) quantum
interrogation and no cooperation, (d) classical interrogation and classical
cooperation, (e) classical interrogation and no cooperation (cf. text). The
dotted base line represents the fundamental bound arising from the finite width
of the clock atoms transition 
[compare \refeq{eq:ADEV2}]. This optimal stability can be attained only via cooperation between the
nodes.
The quantum clock network (a) represents the optimal form of cooperation, and reaches this boundary faster than any other
operational mode. Parameters are $N=1000$, $K=10$, $\gamma_i=10^{-4}\gamma_{LO}$.
 }
\end{figure}

First, we analyze the stability of the proposed quantum clock network,
corresponding to the case of quantum interrogation and cooperation (curve a in
\reffig{fig:comp}). In this case, the analysis resulting in \refeq{eq2}
suggests that near Heisenberg-limited  scaling with a total atom number can be achieved
for the entangled clock network. 
In particular, for a given total particle number $N$ and for averaging times
shorter than the timescale set by individual qubit noise $\tau < 1/(\gamma_i N)$
(where $\gamma_i$ is the atomic linewidth), the network operation achieves a
Heisenberg-limited Allan deviation (ADEV) of the COM laser mode
\begin{align}
\label{eq:ADEV1}
 	\sigma_y(\tau) =  \frac{1}{\omega_0 \sqrt{n_0}2^{M}K} \frac{1}{\tau}
	\sim \frac{ \sqrt{\textrm{log}(N)}}{\omega_0 N} \frac{1}{\tau},
\end{align}
up to small numerical corrections.
Here, the number of GHZ copies per group $n_0\sim \textrm{log}(N)$ ($N\approx n_0
2^{M+1} K $) is found after optimization  [cf. SI], and
gives rise to a logarithmic correction in the total particle number.
The $1/\tau$ scaling results from the effective cancellation of the low
frequency part of the laser noise spectrum, achieved
by the cascaded protocol described above, possibly in combination with additional stages of uncorrelated interrogations using varying Ramsey times \cite{Rosenband:2013vp,Borregaard2013}, see \cite{Kessler2013}.
This allows the cycle time $T$ (which is assumed to be equal to
the interrogation time) to be extended to the total available measurement time
$\tau$.


Eventually, for large averaging times $\tau > 1/(\gamma_i N)$ the
Ramsey time becomes fundamentally limited by individual
noise processes ($T\leq 1/(\gamma_i N)$).
As a result, the $1/N$ scaling breaks down, and the ADEV returns to the square
root scaling with both the employed particle number and averaging time,
\begin{align}
\label{eq:ADEV2}
	\sigma_y(\tau) \sim \frac{ 1}{\omega_0 \sqrt N}
	\sqrt{\frac{\gamma_i}{\tau}},
\end{align}
up to constant numerical factors.  \refeq{eq:ADEV2} results from fundamental
quantum metrological bounds \cite{Escher:2011fn}, and represents the best
conceivable clock stability in the presence of individual particle decoherence
which, in a network, can only be achieved via cooperation. Independently
operating a clock, in contrast, can only achieve a stability scaling with the
local number of atoms, i.e. $\sigma_y(\tau)\propto \sqrt{K/N}$.

 \reffig{fig:comp} illustrates the comparison of entangled clock network with
other approaches.
A network in which the $K$ nodes cooperate classically (curve b in
Fig.~\ref{fig:comp}), by locally measuring the individual phase deviation
$\phi_j$, and combining the outcomes via classical channels, outperforms
individually operated clocks (curve c) by a factor of $\sqrt K$ 
(for both cases, assuming optimal
quantum interrogation for individual nodes
\cite{Kessler2013,Borregaard2013Squeezing}).
The quantum network protocol (curve a)
increases this cooperative advantage by an additional factor of $\sqrt K$ for
short averaging times, reaching Heisenberg-limit. The ADEV
converges to the fundamental bound [\refeq{eq:ADEV2}] $K$ times faster compared
to the case of classical cooperation (curve b).
Although an optimal, classical, local protocol
(e.g. \cite{Rosenband:2013vp, Borregaard2013}), combined with classical
cooperation (curve d), eventually reaches the same
bound [\refeq{eq:ADEV2}],
this approach is atom-shot noise limited, and hence its stability is reduced by
a factor of  $\sqrt N$ for short averaging times [compare \refeq{eq:ADEV1}]
compared to the quantum network protocol.  Hence, the optimal stability
[\refeq{eq:ADEV2}] is reached at averaging times that are $N$ times longer
than for the proposed quantum network. Naturally, all of the above approaches
are superior to a classical scheme without cooperation (curve e).

As a specific example, we first consider ion clocks that can currently achieve a
stability of $2.8\times 10^{-15}$ after $1~\text{s}$ of averaging time
\cite{Chou2010}. The entangled states of up to 14 ions has already been
demonstrated \cite{Monz2011} as was the entanglement of remote ions
\cite{Maunz2007}. We consider a network of ten clocks, each containing ten ions.
Using $\text{Al}^+$ ($\omega_0 = 2\pi \times 1121~\text{THz}$, $\gamma_i = 2\pi
\times 8~\text{mHz}$), we find that the quantum cooperative protocol can reach
$4\times 10^{-17}$ fractional frequency uncertainty after $1~\text{s}$. Even
more pronounced improvement could potentially be achieved using
e.g. $\text{Yb}^+$ ions, due to the long coherence time ($2.2\times
10^4~\text{s}$) of its octupole clock transition.

The quantum gain could be even more pronounced for neutral atomic clocks. For a
network consisting of ten clocks similar to the one operated in JILA
\cite{Nicholson2012}, each containing 1000 neutral atoms with central frequency
$\omega_0 = 2\pi\times 429~\text{THz}$ and linewidth $\gamma_i = 2\pi \times
1~\text{mHz}$,  the quantum cooperative scheme can achieve a stability of $\sim
2\times 10^{-18}$ after 1s averaging, and is an order of magnitude
better than the best classical cooperative scheme.  Future advances,
allowing to employ clock transitions with linewidths of a few tens of
$\mu\text{Hz}$ (such as erbium), could possibly allow for further
improvement, achieving fractional frequency uncertainty beyond $10^{-20}$
after $\tau \sim 100~\text{s}$. This level of stability is in the same order of
magnitude then the required sensitivity to successfully use the network as a
gravitational interferometer \cite{Schiller2008}.


\section*{Security}
\label{sec:Security}
A network with such precise time-keeping capabilities can be subject to both internal
and external attacks. Effectively countering them is crucial to establish a
reliable ground for cooperation. We consider the network secure if the
implemented countermeasures can prevent external parties from benefiting from
the network (eavesdropping), as well as effectively detect any malicious activities of any of the members (sabotage).
%

\begin{figure}
\centering
\includegraphics[width=0.48\textwidth]{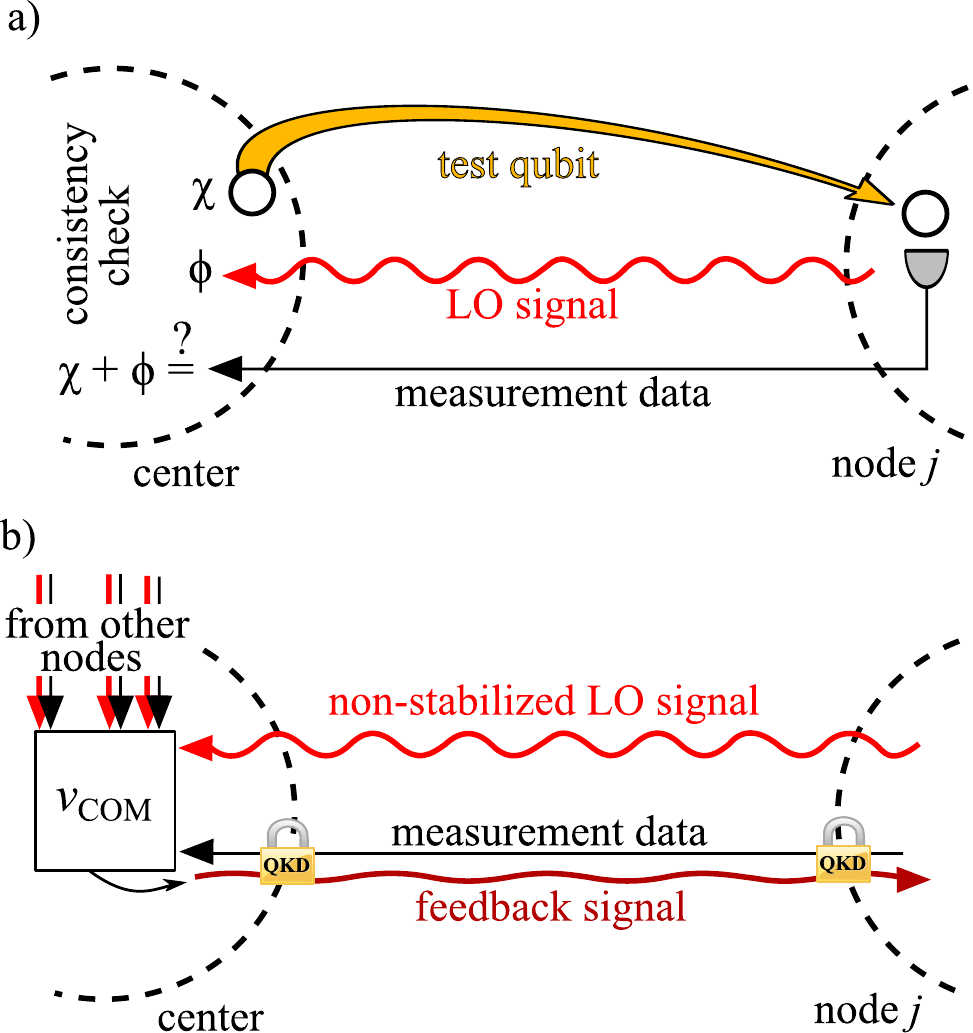}
\caption{
\label{fig:security}
Schematics of security countermeasures.
a) The center node can choose to test any node $j$ by teleporting a disentangled
qubit with a certain phase rotation. A properly operating node creates a local
GHZ state $[\ket{\mathbf{0}} + e^{i\chi}\ket{\mathbf{1}}]/\sqrt{2}$ from the
sent qubit,
 measures the parity of the GHZ state, and sends it to the center. The
measured parity holds information on the phase $\phi' = \chi + \phi$, where
$\phi$ is the accumulated phase of the LO at the node. The center verifies
$\phi$ by comparing it with the classically determined  phase of the sent LO
signal with respect to the COM signal. \\
b) Eavesdropping can be prevented by prescribing that only the non-stabilized LO
signals are sent through  classical channels and encoding the radio frequency
feedback signal with phase  modulation according to a shared secret key.}
\end{figure}

Sabotage describes the situation where one of the nodes -- intended or
unintended -- operates in a damaging manner. For example, one node 
could try sending false LO frequencies or wrong measurement bits in the hope of
corrupting the collective measurement outcomes. In order to detect such
malicious participants, the central node can occasionally perform assessment tests of
the different nodes by teleporting an uncorrelated qubit state $[\ket{0} +
e^{i\chi}\ket{1}]/\sqrt{2}$, where $\chi$ is a
 randomly chosen phase known only to the center. By checking for statistical
discrepancies between the measurement results and the detuning of the LO signal
sent by the node under scrutiny, the center can rapidly and reliably determine whether the particular
node is operating properly (See \reffig{fig:security}a and Supplementary
Information). 

Eavesdropping, i.e., the unauthorized attempt to access the stabilized
$\nu_\text{COM}$ frequency, can be prevented by encoding the classical channels,
over which the center and the nodes exchange feedback signals, using quantum key
distribution protocols \cite{Gisin2002}. Our protocol can keep the stabilized
signal hidden from outsiders by mixing the feedback signal with the LO signal at each node only
after the non-stabilized LO has been sent to the center (see
\reffig{fig:security}b and SI). As a result, even if  all LO signals are intercepted, the eavesdropper is able to access
only the non-stabilized COM signal. Furthermore, the center exclusively can decode the
measurement results sent by the individual nodes using its own measurement outcomes as mentioned above. 
As a result, the stabilized COM signal remains
accessible exclusively to parties involved in the collaboration.

Finally, we note that a distributed operation offers significant security
advantages over an alternative approach of having all resources combined in one
place from where the signal is distributed. In case of a physical attack of the
network, disabling the center or the communication links, the nodes can fall
back to an independent clock operation using their local resources.

\section*{Outlook}



One of the advantages of the proposed quantum clock network involves its ability
to maintain and synchronize the time standards across multiple parties in the
real-time. 
Unlike the current world time
standard, where the individual signals from different clocks are averaged and
communicated with a time delay (a so called paper clock), in our quantum clock
network all participants have access to the ultra-stable signal at any time.
Furthermore, by having full access to their local clocks the different parties keep their full
sovereignty and ensure security, as opposed to a joint operation of a single
clock.

Realization of the full-scale network of the type described here will require a
number of technological advances in both metrology and experimental quantum
information science. 
The remote entanglement can be implemented by using recently demonstrated
techniques for individual atom-photon entanglement \cite{Olmschenk2009,
Chou2007, Togan2010, Bernien2013, Riste2013}. Since the teleportation protocol
requires quantum links capable of sharing EPR pairs with sufficiently high
repetition rate and fidelity, entanglement purification \cite{Dur1999} and
quantum repeater techniques \cite{Duan2001} will likely be required. In
practice, qubits used for entanglement distribution may not be ideal for clocks.
 However, as noted previously  remote entanglement does not need to involve
coherent qubits at optical frequencies (e.g., polarization entanglement can be
used). In such a case,   the use of hybrid approaches, combining  different
systems for entanglement and local clock operations, may be warranted.
It might also be interesting to explore if high-fidelity entangled EPR pairs can
be used to create remote entangled states of spin-squeezed type \cite{Leroux2010,
Sherson2006, Ma2012}, possibly by following the approach for cat state
preparation in remote optical cavities \cite{Haroche_book_2006}, or using local,
collective interactions and repetitive teleportation \cite{Kessler2013b,
Jia2004, Takei2005, Lee2011, Andersen2013}.
In addition, while space-based communication networks will be capable
of maintaining optical phase coherence for the links between clocks, we note
that establishing ground-space coherent optical links remains a technical
challenge and requires an intense research effort which has recently started
\cite{Djerroud2010}.
Finally, if the entire network is spanned by satellites in space, the on-board
local oscillators can further benefit from the much lower noise level compared
to ground-based clocks.

If realized, such a quantum network of clocks can have important
scientific, technological, and social consequences. Besides creating a world platform
for time and frequency metrology, such a network may find important applications
to a range of technological advances for earth science \cite{Tapley2005} and to
the test and search for the fundamental laws of nature, including
relativity and the connection between quantum and gravitational physics
\cite{Abramovici1992, Seidel2007, Schiller2008, Wolf2008}.

%

\appendix 
\section*{Supplementary Information}

\section{GHZ cascade in a network of $K$ clocks}
Here, we discuss the details of using quantum correlated
states constructed out of $N' = Kn$ qubits, equally distributed among $K$
clocks, namely the GHZ state of the form
\bel
	\label{eq:GHZ}
	[\ket{00\dots 0} + e^{i\chi} \ket{11\dots 1}]/\sqrt{2},
\eel
where $\ket{qq\dots q} = \ket{q}^{\otimes N'}$, $q\in\{0,1\}$. Entanglement has
two effects here: First, it makes the phase of such a GHZ state, $\chi$,
sensitive to the accumulated phase of the \emph{center-of-mass} of all the $K$
independent local oscillators, (each located at one of the clocks)
$\Phi_\text{LO} =
\sum_{j=1}^K \Phi^{(j)} /K$, where $\Phi^{(j)} = \intop_0^T\d{t}
(\omega^{(j)}(t) - \omega_0)$ is the accumulated phase of the LO at clock $j$,
during the interrogation time $T$, here $\omega^{(j)}(t)$ is the instantaneous
frequency of the LO, while $\omega_0$ is the transition frequency of the clock
qubit. Second, it increases the sensitivity, due to quantum enhancement:
\bal
	&&\left(\prod_{j}^{K}\prod_i^{N'/K} \hat U_{i,j}\right) \left[\ket{\mathbf{0}}
	+ e^{i\chi}
	\ket{\mathbf{1}}\right]/\sqrt{2}=\nonumber
	\\
	\label{eq:GHZ_evolved}
	&&	\qquad\qquad\qquad = 
	[\ket{\mathbf{0}} + e^{i(\chi + N'\Phi_\text{COM})} \ket{\mathbf{1}}]/\sqrt{2},
\eal
where $\hat U_{i,j} = \ket{0}\bra{0} + e^{i\Phi^{(j)}}\ket{1}\bra{1}$ is the
time evolution operator during the interrogation time, acting on the $i$th qubit
at clock $j$, and $\ket{\mathbf{0}}$ and $\ket{\mathbf{1}}$ are product states of all qubits being in $\ket{0}$ or $\ket{1}$, respectively. 

\subsection{Parity measurement}
By setting the initial phase of the GHZ state, $\chi$, to $0$ and $\pi/2$ in
two parallel instances, we effectively measure the real and imaginary part of
$e^{iN'\Phi_\text{COM}}$, and thus get an estimate on the value of
$N'\Phi_\text{COM}$ up to $2\pi$ phase shifts. The most cost-effective way to
do this is to measure all qubits in the local $x$-basis. In this basis, the
state \refeq{eq:GHZ_evolved} is written as
\bel
	\frac{1}{\sqrt{2}}\left[\left(\frac{\ket{+} -
	\ket{-}}{\sqrt{2}}\right)^{\otimes N'} +  e^{i\phi}\left(\frac{\ket{+}
	+\ket{-}}{\sqrt{2}}\right)^{\otimes N'}\right],
\eel
where $\phi = \chi + N'\Phi_\text{COM}$, and $\ket{\pm} = \frac{\ket{0} \pm
\ket{1}}{\sqrt{2}}$. The above state can be expanded in a sum:
\bel
	\frac{1}{2^{(N'+1)/2}}\sum_{\mathbf{q}\in\{+,-\}^{\times N'}}
	\left[ \left(\prod_{j=1}^{N'}q_j\right) + e^{i\phi}\right]
	\ket{q_1,q_2,\dots q_{N'}},
\eel
where we labeled all qubits with $k\in\{1,2,\dots N'\}$, irrespective of which
clock they belong to.
The probability of a certain outcome $\mathbf{q} = (q_1, q_2, \dots
q_{N'})$, $(q_j \in\{+,-\})$, is
\bel
	\PP(\mathbf{q}) = \frac{1}{2^{N' +1}} |1+ p(\mathbf{q})e^{i\phi}|^2,
\eel
where $p(\mathbf{q}) = \prod_{j=1}^{N'}q_j$ is the parity of the sum of all
measurement bits. Now, the clocks send their measurement bits to the center
node, which evaluates $p$.
This parity is the global observable that is sensitive to the accumulated phase,
since its distribution is
\bel
	\PP(p=\pm) = \frac{1\pm \cos(\phi)}{2}.
\eel
The above procedure is identical to the parity measurement scheme described in
\cite{Bollinger1996}.

\subsection{Cascaded GHZ scheme}
\begin{figure}
\centering 
\includegraphics[width=0.48\textwidth]{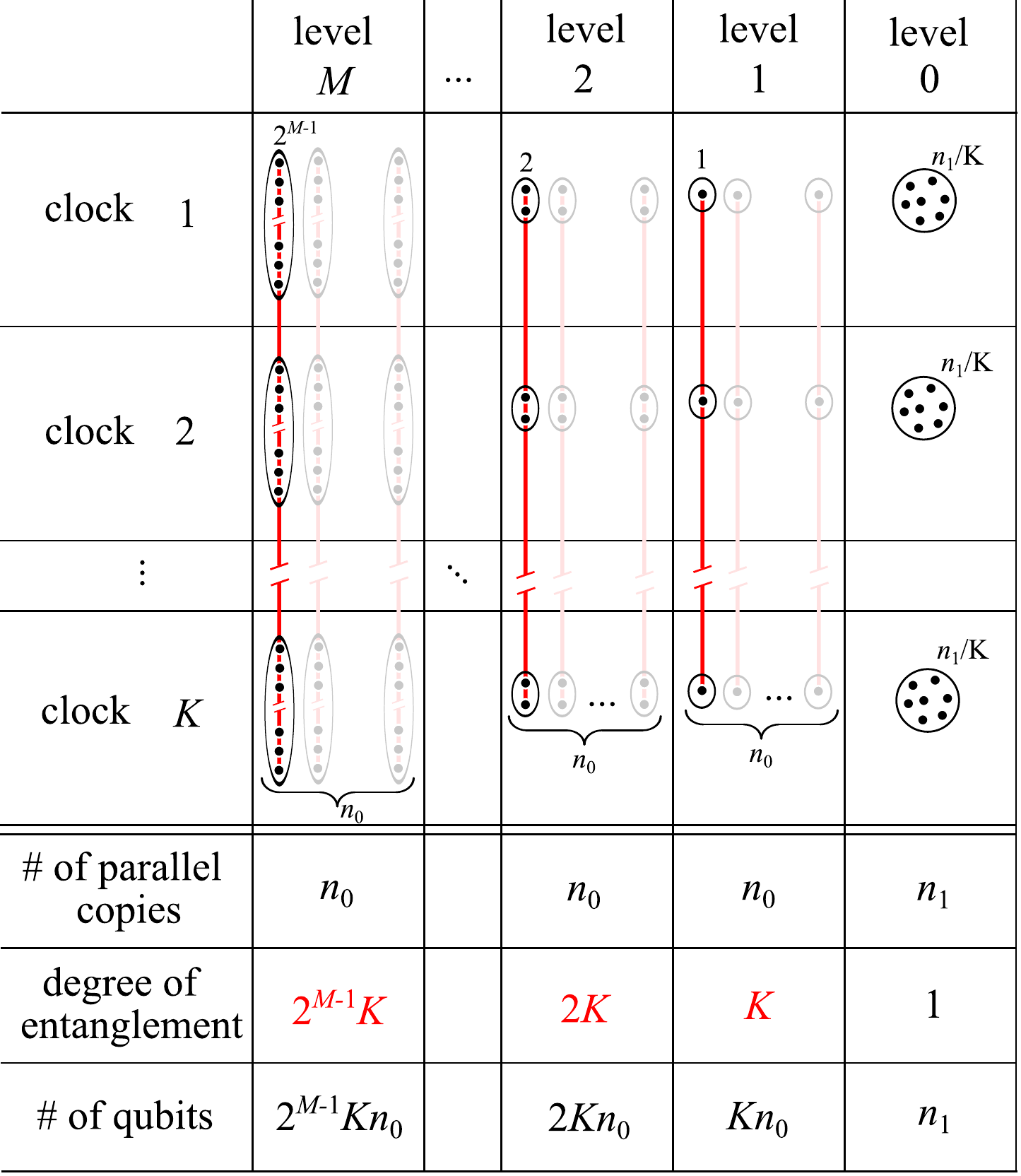}  
\caption{
\label{fig:GHZ_cascade}
GHZ cascade protocol for $K$ clocks. Each allocates qubits for different levels
of the protocol: In level 0, $n_1/K$ qubits are put into an uncorrelated
ensemble. In level $i$, $(i=1,2\dots M)$, each clock allocates $n_0 2^{i-1}$
qubits for creating $n_0$ parallel instances of GHZ states with $2^{i-1}K$
entangled qubits. Due to the exponential scaling of the degree of entanglement, most of
the total available qubits are used in higher levels of the cascade. This is a
necessary condition to achieve Heisenberg scaling, up to logarithmic factors.}
\end{figure}

Provided with $N$ qubits distributed equally among $K$ clocks, we imagine
that each clock separates its qubits into $M+1$ different groups. The $0$th
group contains $n_1/K$ uncorrelated qubits, and the $i$th group ($i=1,2\dots M$)
contains $n_0$ independent instances of $2^{i-1}$ qubits that are entangled with
the other groups of $2^{i-1}$ qubits in each clock. In other words, there
are $n_0$ independent copies of GHZ states with a total of $2^{i-1} K$ qubits
entangled on the $i$th level of the cascade ($i\geq 1$) (See
\reffig{fig:GHZ_cascade}). This way the total number of qubits can be written as
\bel
	N = n_1 + n_0 \sum_{i=1}^{M} 2^{i-1}K \approx n_0 2^{M}K
\eel
where we assumed $n_1 \ll N$. 
  
The purpose of this cascaded scheme is to directly assess the digits $Y_1$ and
$\{Z_j\,:\, j=2,3,\dots\}$ in the binary fraction representation of the phase
\bel
	\Phi_\text{LO}\mod[-\pi,\pi] = \frac{2\pi}{K}\left[Y_1 +\sum_{i=1}^\infty
	Z_{i+1}/2^i\right] - \pi,
\eel
where $\mod [-\pi, \pi] = (x+\pi) \mod 2\pi - \pi$, $Y_1\in\{0,1,2\dots K-1\}$
and $Z_i\in\{0,1\}$. The $0$th level of the
cascade estimates $\Phi_0 =
\sum_{j=1}^K \left(\Phi^{(j)}\mod[-\pi,\pi]\right)/K$, and every $i$th
level after that estimates $\Phi_i = K2^{i-1}\Phi_\text{LO} \mod [-\pi,\pi]$. From these estimates one can determine the digits,
%
%
\bal
	Y_1 &=& \left[K(\Phi_0 + \pi) - (\Phi_1 + \pi)\right]/(2\pi),
	\\
	Z_i &=& \left[2(\Phi_{i-1} + \pi) - (\Phi_i + \pi)\right]/(2\pi),
\eal 
for $i=2,3, \dots M$. 

The last group ($i=M$) contains GHZ states with the most entangled qubits.
These are the ones with the fastest evolving phase, and therefore they provide
the best resolution on $\Phi_\text{LO}$. Since
there are $n_0$ independent instances, their phase $\Phi_{M} = 2\pi \sum_{i=1}^\infty Z_{M+i}/2^i$ is known
up to the uncertainty, $\ev{\Delta\Phi^2_{M}}_\text{pr} =
\frac{1}{n_0}$, 

Assuming that all lower digits $\{Y_1,Z_j|j=2\hdots M \}$ have been determined correctly, this results in the total measurement uncertainty for $\Phi_\text{LO}$:
\bel
	\label{eq:projection_GHZ 2}
	\ev{\Delta\Phi^2_\text{LO}}_\text{pr} =
	\frac{\ev{\Delta\Phi^2_{M}}_\text{pr}} {(2^{M-1}K)^2} = \frac{4n_0}{N^2},
\eel
where, for the moment, we neglected individual qubit noise and assumed $\Phi_\text{LO}\in[-\pi,\pi]$.
However, in general, the estimation of the lower digits will not be perfect. In the following Section we investigate the effect of these \textit{rounding errors} on the final measurement accuracy. From this analysis we find the optimal number of copies $n_0$ and $n_1$. 

\subsection{Rounding errors}
Whenever $|\Phi_0^\text{est} - \Phi_0| > \pi/K$, or
$|\Phi_i^\text{est} - \Phi_i| > \pi/2$ (for $i \geq 1$), we make a mistake by
under- or overestimating the number of phase slips $Y_1$ or $Z_{i+1}$,
respectively. To minimize the effect of this error, we need to optimize how the
total of $N$ qubits are distributed among the  various levels of the cascade. In
other words we need to find $n_{0,\text{opt}}$ and $n_{1,\text{opt}}$.

The probability that a rounding error occurs during the estimation of $Z_{i+1}$
is
\bel
	\PP_{i,\text{re}} = 2\intop_{\pi/D}^\infty \d{\phi}
	\rho_i(\phi - \Phi_i) \leq 2\intop_{\pi/D}^\infty \d{\phi} \frac{1}{s_i^3}
	\exp\left[-\frac{\phi^2}{2s_i^2}\right]
\eel
where $\phi = \Phi_i^{\text{est}} - \Phi_i$, and $\rho_i$ is the conditional
density function of $\Phi_i^\text{est}$ for a given real $\Phi_i$, and $s_i^2 =
\text{Var}(\Phi_i^\text{est} - \Phi_i) = 1/n_0$ for $i\geq 1$, and $s_0^2 =
\ev{\Delta\Phi_0^2}_\text{pr} = \frac{1}{K^2}\sum_{j=1}^K
\ev{(\Delta\Phi^{(j)})^2}_\text{pr} = 1/n_1$, since
$\ev{(\Delta\Phi^{(j)})^2}_\text{pr} = \frac{K}{n_1}$ for all $j$.
The upper bound for $\rho_i$ is obtained 
by using the following upper bound for any binomial
distribution: ${m\choose k}p^k(1-p)^{m-k} \leq
\exp\left[-n\left(\frac{k}{n}-p\right)^2\right]$. (For details, see
Supplementary Materials of \cite{Kessler2013}.)  The resulting probabilities, after
dropping the higher order terms in the asymptotic expansions, are
\bal
	\PP_{0,\text{re}} &\approx& \frac{2K}{\pi} n_1^{1/2}
	\exp\left[-\frac{n_1\pi^2}{2K^2}\right]
	\\
	\PP_{i,\text{re}} &\approx& \frac{4}{\pi} n_0^{1/2}
	\exp\left[-\frac{n_0\pi^2}{8}\right]\qquad (i\geq 1)
\eal

The phase shift imposed on the estimate of $\Phi_\text{LO}$ by a
manifested rounding error of $Y_1$ is $2\pi/K$ and of $Z_{i}$ is $2\pi/
(K 2^{i-1})$, for $i=2,3\dots M$.
This
results in the total variance contribution,
\bal
	&&\ev{\Delta\Phi^2_\text{LO}}_\text{re}  =\nonumber\\
	&&\quad 
	=\left(\frac{2\pi}{K}\right)^2\left[\PP_{0,\text{re}} +
	\sum_{i=2}^{M}\PP_{i-1,\text{re}} (2^{-i+1})^2\right]
	\\
	&&\quad
	\approx\left(\frac{2\pi}{K}\right)^2
	\left[\PP_{0,\text{re}} +
	\frac{1}{3}\PP_{i-1,\text{re}}\right].
\eal
We simplify this expression by choosing $n_1$ so that  $\PP_{0,\text{re}}
\approx \frac{2}{3} \PP_{i,\text{re}}$:
\bel
	n_1 = \alpha K^2 n_0,
\eel
where $\alpha \approx \text{max}\left\{1\;,\; \frac{2}{\pi^2 n_0}
\log\left(3K^2\frac{\sqrt{8}}{\pi n_0^{1/2}}\right)\right\} \ll n_0,
K$.
With this choice, we can write the rounding error contribution as
\bel
	\label{eq:Rounding_GHZ 2}
	\ev{\Delta\Phi^2_\text{LO}}_\text{re} \approx \frac{16\pi}{K^2}
	n_0^{1/2} \exp\left[-\frac{n_0 \pi^2}{8}\right].
\eel
We note that the amount of extra resources needed for the 0th level, is
marginally small, since the total qubit number can be expressed as
\bel
	N = n_1 + n_0 K \sum_{i=1}^M 2^{i-1} = n_0 K(\alpha K^2 + 2^{M} -2) \approx
	n_0 K2^{M},
\eel
under the assumption $K \ll 2^{M/2}$.

By adding the two error contributions from \refeq{eq:projection_GHZ 2} and
\refeq{eq:Slip_GHZ 2}, we obtain the corresponding Allan-variance,
\bal
	\sigma_y^2(\tau) &=& \frac{1}{\omega_0^2 \tau T} \ev{\Delta\Phi_\text{LO}^2}   
	=:
	\frac{1}{\omega_0^2\tau}\left[\Gamma_1 + \Gamma_2\right] =\\	
	&=& \frac{1}{\omega_0^2\tau}\left[\frac{4n_0}{N^2 T} +
	\frac{16\pi}{K^2 T} n_0^{1/2}\exp\left[-\frac{n_0\pi^2}{8}\right] 
	\right]
\eal
Now, let us find the optimal value of $n_0$. We write $\Gamma_1 + \Gamma_2$,
using the new variable $x = \frac{8}{\pi^2}\frac{1}{n_0}$, as
\bel
	\Gamma_1 + \Gamma_2 = \frac{4}{T}\left(\frac{8}{\pi^2}\frac{1}{x N^2} +
	\frac{\sqrt{32}}{K^2}\frac{1}{x^{1/2}}\exp\left[-\frac{1}{x}\right]\right).
\eel
Taking the derivative with respect to $x$ and equating it with 0, while using
the assumption $x \ll 1$ results in $\Gamma_2 \approx
x_\text{opt}\Gamma_1 \ll \Gamma_1$, which can be written as the
following transcendental equation for the optimal value, $x_\text{opt}$,
\bel
	x_\text{opt}^{1/2} \approx 
	\frac{\pi^2N^2}{\sqrt{8}K^2}\exp\left[-\frac{1}{x_\text{opt}}\right].
\eel
The general solution of any equation of the form $x^\nu = A \exp[-1/x]$, in the
limit of $A \gg 1$ and $x \ll 1$, is $x = [\log(A)]^{-1}$ . (For details, see
the Supplementary Materials of \cite{Kessler2013}.) Using this result we can write
\bal
	x_\text{opt} &\approx&
	\left[\log\left(\frac{\pi^2}{\sqrt{8}}\frac{N^2}{K^2}\right)\right]^{-1} \sim
	[2\log(N/K)]^{-1}
	\\
	n_{0,\text{opt}} &\approx&
	\frac{8}{\pi^2}\frac{1}{x_\text{opt}} \sim \left(\frac{4}{\pi}\right)^2
	\log\left(N/K\right).
\eal
For the realistic case of $N/K \gg 1$, indeed $x_\text{opt} \ll 1$, and the
corresponding minimal value of $\Gamma_1 + \Gamma_2$ is
\bel
	\label{eq:Gamma_1 + Gamma_2 min}
	[\Gamma_1 + \Gamma_2]_\text{min} \approx \Gamma_1(x_\text{opt}) =
	\left(\frac{8}{\pi}\right)^2\frac{\log(N/K)}{N^2 T}.
\eel
This result indicates that, in terms of qubit number, only a
logarithmic extra cost is required to achieve the Heisenberg limit. 

\subsection{Phase slip errors}
Although the cascade is designed to detect phase slips of all levels $i=1,2\dots
M$, a possible phase wrap of level $i=0$ remains undetected. Since the
qubits at different clocks are interrogated independently on the $0$th
level, each of them estimates the phase of the corresponding LO, $\Phi_0^{(j)}$
($j=1,2, \dots K$), and not $\Phi_\text{LO}$. The probability of $\Phi_0^{(j)}$
falling outside the interval $[-\pi,\pi]$ at least once during the total
measurement time $\tau$ is
\bal
	\PP_{j,\text{slip}} &=& 2\frac{\tau}{T} \intop_\pi^{\infty}\d{\phi}
	\frac{1}{\sqrt{2\pi\gamma_\text{LO}
	T}}\exp\left[-\frac{\phi^2}{2\gamma_\text{LO}T}\right] \approx 
	\nonumber\\
	&\approx&
	\frac{\tau}{T}\frac{\sqrt{2}}{\pi^{3/2}}\sqrt{\gamma_\text{LO}T}
	\exp\left[-\frac{\pi^2}{2\gamma_\text{LO}T}\right],
\eal
where $\gamma_\text{LO}$ is the linewidth of the local oscillator at clock $j$,
corresponding to a white noise spectrum, resulting in a constant phase 
diffusion over the interrogation time $T$, (which assumed to be
approximately equal to the cycle time). The approximate form above is obtained
by neglecting the higher order terms in the asymptotic series expansion under the
assumption $\gamma_\text{LO}T \ll 1$. Once such a phase slip happens, it
introduces a $2\pi$ phase shift in $\Phi_0^{(j)}$, and therefore contributes
to its overall uncertainty with $\ev{(\Delta\Phi_0^{(j)})^2} = (2\pi)^2
\PP_{j,\text{slip}}$. Physically $\Phi_0$ is the phase of the COM signal, that
the center can obtain after averaging the frequencies of all $K$ local
oscillators with equal weights, $\Phi_0 = \Phi_\text{COM} =
\sum_{j=1}^K \Phi_{0}^{(j)}/K$, therefore
$
	\ev{\Delta\Phi_0^2} = \frac{1}{K^2}\sum_{j=1}^K \ev{(\Delta\Phi_0^{(j)})^2} =
	\frac{1}{K}\ev{(\Delta\Phi_0^{(j)})^2},
$
where we assumed that the LOs are independent but they have the same linewidth,
$\gamma_\text{LO}$. Since $\Phi_0 = \Phi_\text{LO}$, the above means the 
following variance contribution
\bel
	\label{eq:Slip_GHZ 2}
	\ev{\Delta\Phi^2_\text{LO}}_\text{slip} =
	\sqrt{32\pi} \frac{\tau \gamma_\text{LO}^{1/2}}{T^{1/2} K}
	\exp\left[-\frac{\pi^2}{2\gamma_\text{LO}T}\right].
\eel
After adding this error to the previously minimized projection and rounding
error terms (from \refeq{eq:Gamma_1 + Gamma_2 min}), we obtain the corresponding
Allan-variance, $\sigma_y^2(\tau) =\frac{1}{\omega_0^2\tau}\left([\Gamma_1 + \Gamma_2]_\text{min} +
\Gamma_3\right)$, where
\bal
	&&[\Gamma_1 + \Gamma_2]_\text{min} + \Gamma_3 =
	\\ 
	&&\quad=\left(\frac{8}{\pi}\right)^2
	\frac{\log(N/K)}{N^2}\frac{2\gamma_\text{LO}}{\pi^2}\frac{1}{y} +
	\frac{16}{\pi^{5/2}}\frac{\tau\gamma_\text{LO}^2}{K}\frac{1}{y^{3/2}}
	\exp\left[-\frac{1}{y}\right],
	\nonumber
\eal
using the variable $y = \frac{2}{\pi^2}\gamma_\text{LO}T$. 

Now, let us find the optimal
Ramsey time $T_\text{opt}$, under the assumption that $\tau$ is sufficiently
long.
After taking the
derivative with respect to $y$ and equating it with zero, the assumption $y_\text{opt}\ll
1$ results in the $\Gamma_3 \approx y_\text{opt}[\Gamma_1 +
\Gamma_2]_\text{min} \ll [\Gamma_1 +
\Gamma_2]_\text{min}$ which can be written as the following transcendental
equation,
\bel
	y_\text{opt}^{3/2} \approx
	\frac{\pi^{3/2}}{8}\frac{\tau\gamma_\text{LO}}{K}\frac{N^2}{\log(N/K)}
	\exp\left[-\frac{1}{y}\right].
\eel
The asymptotic solution in case of $y_\text{opt}\ll 1$ is (see Supplementary
of \cite{Kessler2013})
\bal
	y_\text{opt} &\approx&
	\left[\log\left(\frac{\pi^{3/2}}{8}\frac{\tau
	\gamma_\text{LO}}{K}\frac{N^2}{\log(N/K)}\right)\right]^{-1},
	\\
	\label{eq:T_op_GHZ 2}
	T_\text{opt} &\approx& \frac{\pi^2}{2}\frac{y_\text{opt}}{\gamma_\text{LO}}
	\sim
	\frac{\pi^2}{2\gamma_\text{LO}}\left[\log(\tau\gamma_\text{LO}N^2/K)\right]^{-1}
\eal
in the realistic limit of $\gamma_\text{LO}\tau N^2/K \gg 1$. The corresponding
minimal Allan-variance is
\bel
	\label{eq:long_tau}
	\sigma_y^2(\tau) = \frac{1}{\omega_0^2\tau}\Big[[\Gamma_1 +
	\Gamma_2]_\text{min} + \Gamma_3\Big]_\text{min} \approx
	\frac{1}{\omega_0^2}\frac{L\gamma_\text{LO}}{N^2 \tau},
\eel
where $L =
\frac{128}{\pi^4}\log(N/K)\log(\tau\gamma_\text{LO}N^2/K)$.

For short $\tau$ averaging times, the optimal Ramsey time is
$T_\text{opt}=\tau$, instead of \refeq{eq:T_op_GHZ 2}. This makes $\Gamma_3$
negligible compared to $[\Gamma_1 + \Gamma_2]_\text{min}$, resulting in a
$1/\tau^2$ scaling:
\bal
	\label{eq:short_tau}
	\sigma_y^2(\tau)=\frac{1}{\omega_0^2\tau}[\Gamma_1 +
	\Gamma_2]_\text{min}^{T=\tau} = \frac{1}{\omega_0^2}\frac{L'}{N^2 \tau^2}.
\eal
where $L' = \left(\frac{8}{\pi}\right)^2 \log(N/K)$. This scaling is more
favorable, but it continues to higher $\tau$ values only up to $\tau \sim
\gamma_\text{LO}^{-1}$, where it switches to the $1/\tau$ behavior according to
\refeq{eq:long_tau}.

\subsection{Pre-narrowing the linewidth}
We can minimize the limiting effect of $\gamma_\text{LO}$ by narrowing the
effective linewidth of the local oscillators beforehand. We imagine using $N\s$
qubits to locally pre-narrow
the linewidth of all LOs down to an effective linewidth $\gamma_\text{eff} \sim
\gamma_\text{ind}N$, before using the rest $N-N\s$ qubits in the GHZ cascade.
This $\gamma_\text{eff} \ll \gamma_\text{LO}$ allows the optimal Ramsey time
going above the previous limit, set by $\sim \gamma_\text{LO}^{-1}$ in
\refeq{eq:T_op_GHZ 2}.
This step-by-step linewidth narrowing procedure, using uncorrelated ensembles in
every step, is outlined in \cite{Rosenband:2013vp, Borregaard2013}, and given
detailed analysis in \cite{Kessler2013}. Working under the small $N\s$ assumption,
one can obtain $\gamma_\text{eff}$ as
\bel
	\gamma_\text{eff} \approx 
	\gamma_\text{LO}\left[\frac{2}{\pi^2}\frac{\log(\gamma_\text{LO}\tau
	n)}{n}\right]^{N\s/n},
\eel
where we imagine using $n$ qubits in each narrowing step. We find the optimal
value of $n$ to be
\bel
	n_\text{opt} \approx \frac{2e}{\pi^2}\log(\gamma_\text{LO}\tau),
\eel 
by minimizing $\gamma_\text{eff}$, which yields
\bel
	\label{eq:gamma_m2}
	[\gamma_\text{eff}]_\text{min} \sim \gamma_\text{LO}\exp\left[-\frac{N\s
	\pi^2}{2e\log(\gamma_\text{LO}\tau)}\right].
\eel

For a given $\tau$, we can always imagine carrying out this pre-narrowing, so
that $\gamma_\text{eff} < \tau^{-1}$, and therefore \refeq{eq:short_tau} remains
valid with the substitution $N\mapsto N-N\s$ for $\tau >
\gamma_\text{LO}^{-1}$ as well. The required number of qubits, $N\s$, is
\bel
	N\s \sim
	\frac{2e}{\pi^2}\log(\gamma_\text{LO}\tau)
	\log\left(\frac{\gamma_\text{LO}}{\gamma_\text{ind}N}\right) \ll N.
\eel
due to the exponential dependence in \refeq{eq:gamma_m2}.

\subsection{Individual qubit dephasing noise}
Our scheme, as well as any scheme, is eventually limited by individual qubit
noise. Such a noise dephases GHZ states at an increased rate, compared to
uncorrelated qubits, due to the entanglement, giving the corresponding variance
contribution for the phase of the GHZ states in the $M$th group, 
$\ev{\Delta\Phi^2_{M}}_\text{dephasing} =  \frac{2^{M-1}
K\gamma_\text{ind}T}{n_0}$, after averaging over the $n_0$ independent copies of the GHZ states, each
containing $2^{M-1} K$ entangled qubits. The resulting variance
contribution for $\Phi_\text{LO}$ is
\bel
	\label{eq:Dephasing_GHZ 2}
	\ev{\Delta\Phi^2_\text{LO}}_\text{dephasing} =
	\frac{ \gamma_\text{ind}T}{n_0 2^{M-1}K} = \frac{2\gamma_\text{ind}T}{N}.
\eel
This term represents a noise floor, which we add to
\refeq{eq:short_tau} and obtain our final result for the minimal achievable
Allan-variance,
\bel
	\sigma_y^2(\tau) = \frac{1}{\omega_0^2}\left[\frac{L'}{N^2 \tau^2} +
	\frac{2\gamma_\text{ind}}{N\tau}\right].
\eel

For long $\tau$ times, the ultimate limit, set by the standard quantum limit,
$
	\sigma_y^2(\tau) = \frac{1}{\omega_0^2}
	\frac{\gamma_\text{ind}}{N\tau},
$
can be reached by changing the base of the cascade. Instead of entangling
2-times as many qubits in each level of the cascade than in the previous level,
we imagine changing it to a base number $D$. Carrying out the same calculation
results in our final result for the achievable Allan-variance:
\bel
	\sigma_y^2(\tau) =
	\frac{1}{\omega_0^2}\left[\left(\frac{D}{2}\right)^2\frac{L'}{N^2 \tau^2} +
	\frac{D}{D-1}\frac{\gamma_\text{ind}}{N\tau}\right],
\eel
where $L' = \left(\frac{8}{\pi}\right)^2 \log(N/K)$.
(See Supplementary of \cite{Kessler2013} for details.)
The optimal value of $D$ depends on $\tau$. For small $\tau$, $D_\text{opt} =
2$, however for large $\tau$ one can gain a factor of 2 by choosing $D_\text{opt} = D_\text{max}$. Due to natural
constraints, $D_\text{max} \sim \sqrt{N}$, in which regime, the protocol
consists of only two cascade levels, an uncorrelated 0th level, with $\sim
\sqrt{N}$ qubits and an entangled 1st level with $\sim N$ qubits.

\section{Security countermeasures}

\subsection{Sabotage}
In order to detect sabotage, the center can occasionally perform assessment
tests of the different nodes by teleporting an uncorrelated qubit state $[\ket{0} +
e^{i\chi}\ket{1}]/\sqrt{2}$, where $\chi$ is a
 randomly chosen phase known only to the center.
A properly operating node creates a local GHZ state $[\ket{\mathbf{0}} +
e^{i\chi}\ket{\mathbf{1}}]/\sqrt{2}$ from the sent qubit,
 measures the parity of the GHZ state, and sends it to the center. The
measured parity holds information on the phase $\phi' = \chi + \phi$, where
$\phi$ is the accumulated phase of the LO at the node. Due to the random shift
$\chi$, this appears to be random to the node, and therefore indistinguishable
from the result of a regular (non-testing) cycle. On the other hand, the center
can subtract $\chi$, and recover $\phi$ from the same measurement results. In
the last step, the center verifies $\phi$ by comparing it with the classically
determined phase $\phi_\text{cl}$ of the sent LO signal with respect to the COM
signal. The expected statistical deviation of
$\phi$ from $\phi_\text{cl}$ is $\Delta(\phi -
\phi_\text{cl})\sim\sqrt{\frac{K}{N}}$, while the accuracy of the COM phase
$\Delta(\phi_\text{COM} - T\omega_0)\sim\sqrt{\frac{K}{(K-K_t)N}}$ is much
smaller, where $K_t$ is the number of simultaneously tested nodes. In the likely
case of $K_t \ll K$, this method is precise enough for the center to
discriminate between healthy and unhealthy nodes by setting a acceptance range,
$|\phi - \phi_\text{cl}| \leq \Lambda \sqrt{\frac{K}{N}}$. E.g. the choice of
$\Lambda = 4$ results in a ``$4\sigma$ confidence level'', meaning only 0.0063\%
chance for false positives (healthy node detected as unhealthy), and similarly
small chance for false negatives (unhealthy node being undetected)
$(\sim\Lambda\frac{\Delta \phi'}{2\pi}\propto 1/\sqrt{N}) $ due to the high
precision with which $\phi'$ is measured.
The fact, that the teleported qubit can be measured only once, also prevents the
nodes from discovering that it is being tested.

\subsection{Eavesdropping}
Eavesdroppers would try to intercept the sent LO
signals, and synthesize the stabilized $\nu_\text{COM}$ for themselves. Our
protocol minimizes the attainable information of this strategy by prescribing
that only the \emph{non-stabilized} LO signals are sent through classical
channels. This requires the feedback to be applied to the LO signal after some
of it has been split off by a beam splitter, and the center to integrate the
generated feedback in time. Alternatively, eavesdroppers could try intercepting
the LO signals \emph{and} the feedback signals, and gain access to the same
information, the center has. This can be prevented by encoding the radio
frequency feedback signal with phase modulation according to a shared secret
key. Since such a key can be shared securely with quantum key distribution, this protocol keeps the feedback signal hidden from outsiders. As a
result, even the hardest-working eavesdropper, who intercepts all LO signals, is
able to access only the non-stabilized COM signal, and the stabilized COM signal
remains accessible exclusively to parties involved in the collaboration.

\subsection{Rotating center role}
Since the center works as  a hub for all information, ensuring
its security has the highest priority. In a scenario,  where none of the nodes
can be trusted enough to play the permanent role of the center, a rotating stage
scheme can be used. By passing the role of the center around, the
potential vulnerability of the network due to one untrustworthy site is
substantially lowered. This requires a fully connected
network and a global scheme for assigning the role of the center.

\section{Network operation}

\subsection{Different degree of feedback}
Apart from the full feedback, described in the main text, alternatively, the
center can be operated to provide restricted feedback information to the nodes.
If the center sends the averaged error signal $\tilde \delta_\text{COM}$ only,
the LOs at the nodes will not benefit from the enhanced stability and only the
center can access the stabilized signal. Of course the LO at each node will have
its own local feedback to keep it within a reasonable frequency range around the
clock transition. Such a 'safe' operational mode makes the center node the
only participant having access to the world time signal.

As an intermediate possibility, the center can choose to send regionally
averaged feedback signals $\tilde\delta_\text{COM} + \sum_{j\in R}(\nu_j -
\nu_\text{COM})/|R|$, uniformly for all $j\in R$ nodes, where $R$ is a set of
nodes, ie. a region. Such a feedback scheme creates the incentive of cooperation
for the nodes in region $R$. By properly sharing their LO signals with each
other, the nodes can synthesize the regional COM frequency, $\sum_{j\in
R}(\nu_j)/|R|$, and steer it with the feedback, received from the center.

\subsection{Timing}
Proper timing of local qubit operations is necessary to ensure that every qubit
in the network is subject to the same $T$ free evolution time.
The finite propagation time of light signals introduces delays in the quantum
links and classical channels.
Similarly, during the entangling step, the finite time required to do CNOT
operations make the free evolution start at slightly different times for
different qubits. Since both the initialization and the measurement are local
operations, we can resolve the issue of delay by prescribing that the
measurement of qubit $i_j$ ($i$th qubit at node $j$) takes place exactly $T$
time after its initialization. Occasional waiting times of known length can be
echoed out with a $\pi$-pulse at half time. 

In extreme cases, this might cause
some qubits to be measured before others are initialized. However, this is not a
problem, since the portion of the GHZ state that is alive during the
time in question is constantly accumulating the $\phi_j$ phases from the qubits
it consists of. This results in the phenomenon that the total time of phase accumulation can be much longer than the length of individual
phase accumulations, provided that the said interrogations overlap.

\subsection{More general architectures}
So far, we focused on the simplest network structure with one center
initiating every Ramsey cycle and nodes with equal number of clock qubits.

In a more general setup, node $j$ has $N_j$ clock qubits. If $N_j$ is different
for different $j$, then the nodes will contribute the the global GHZ states
unequally, resulting in entangled states which consists of different $N_j'$
number of qubits from each site $j$. Such a state picks up the phase
\bel
	\Phi = \sum_{j}N'_j \phi_j,
\eel
where $\phi_j$ is the phase of the LO at site $j$ relative to the atomic
frequency. As a result, the clock network measures the following collective LO
frequency
\bel
	\nu_\text{LO} = \frac{\sum_{j}N'_j \nu_j }{ \sum_{j}N'_j}.
\eel
This represents only a different definition of the world time (a weighted
average of the times at the locations of the nodes, instead of a uniform
average), but it does not affect the overall stability.

The initial laser linewidths of the nodes $\gamma_\text{LO}^j$ can also be
different. The stability achievable in this case is bounded by the stability
obtained for a uniform linewidth $\gamma_\text{LO} = \max_j \gamma_\text{LO}^j$.
If linewidths are known, the center can divise the best estimation method
which uses linewidth dependent weights in the LO frequency averaging step. 

Although it is simple to demonstrate the important network operational concepts
with the architecture with one center, this structure is not a necessary. The
quantum channels, connecting different nodes, can form a sparse (but still connected)
graph, and the entanglement global entanglement can still be achieved by
intermediate nodes acting as repeater stations. This way entanglement can be
passed along by these intermediate nodes. Moreover, the center can be eliminated
from the entangling procedure by making the nodes generate local GHZ states, and
connect them with their neighbors by both measuring their shared EPR qubit with
one of the qubits form the local GHZ state in the Bell-basis. After
communicating the measurement result via classical channels, and performing the
required single qubit operations, a global GHZ state is formed.
 
%

 \section*{Acknowledgement}
We are grateful to Till Rosenband and Vladan Vuleti\'{c} for enlightening
discussions.
This work was supported by NSF, CUA, ITAMP, HQOC, JILA PFC, NIST, DARPA QUSAR,
the Alfred P. Sloan Foundation, the Quiness
programs, ARO MURI, and the ERC grant QIOS (grant no. 306576); MB
acknowledges support from NDSEG and NSF GRFP.
It is dedicated to Rainer Blatt and Peter Zoller on the occasion of their 60th
birthday, when initial ideas for this work were formed.
 
 \section*{Competing financial interests}
 The authors declare no competing financial interests.
\end{document}